\documentclass[aip,jcp,preprint,superscriptaddress,floatfix]{revtex4-1}

\usepackage[utf8]{inputenc}
\usepackage{color}
\usepackage{graphicx}
\usepackage{mathtools}
\usepackage{bm}
\usepackage{amsmath,amsfonts,amssymb}

\begin{document}
\title{Error Assessment of Computational Models in Chemistry}

\author{Gregor N. Simm}
\affiliation{%
        ETH Z{\"u}rich, Laboratorium f\"ur Physikalische Chemie, Vladimir-Prelog-Weg~2, 8093 Z\"urich, Switzerland
}%

\author{Jonny Proppe}
\affiliation{%
        ETH Z{\"u}rich, Laboratorium f\"ur Physikalische Chemie, Vladimir-Prelog-Weg~2, 8093 Z\"urich, Switzerland
}%

\author{Markus Reiher}
\email{markus.reiher@phys.chem.ethz.ch}
\affiliation{%
        ETH Z{\"u}rich, Laboratorium f\"ur Physikalische Chemie, Vladimir-Prelog-Weg~2, 8093 Z\"urich, Switzerland
}%

\begin{abstract}
Computational models in chemistry rely on a number of approximations.
The effect of such approximations on observables derived from them is often unpredictable.
Therefore, it is challenging to quantify the uncertainty of a computational result, which, however,
is necessary to assess the suitability of a computational model.
Common performance statistics such as the mean absolute error are prone to failure as they do not distinguish the explainable (systematic) part of the errors from their unexplainable (random) part.
In this paper, we discuss problems and solutions for performance assessment of computational models based on several examples from the quantum chemistry literature.
For this purpose, we elucidate the different sources of uncertainty, the elimination of systematic errors, and the combination of individual uncertainty components to the uncertainty of a prediction.
\end{abstract}

\maketitle

\section{Introduction}
Quantum electrodynamics (QED) allows for the description of all electromagnetic processes occurring between the elementary particles of chemical systems (e.g., molecules).
It is the fundamental theory of chemistry (focusing on the dominant electromagnetic interactions and ignoring the other fundamental forces).
If we were able to solve its equations for some chemical system with arbitrary accuracy, truly predictive results bare of almost all errors would be obtained.
However, for all but the simplest systems, calculations based on QED are unfeasible.
Additional approximations have to be made for the calculation of an observable of interest to be available in reasonable time and with reasonable effort leading to deviations from the fundamental theory of chemistry.
Eventually, the number and types of approximations necessary for a feasible description of molecular systems are vast and diverse.

The precise effect of such approximations (computational models) on observables derived from them is generally unknown and difficult to estimate for arbitrary molecules \cite{Glotzer2009}.
While the procedure of uncertainty quantification for physical measurements is well established \cite{ISOUncertainty1995}, this is not the case for results of computational models (virtual measurements \cite{Irikura2004}).
By the very nature of a deterministic (or fully converged stochastic) calculation, the repetition of such a calculation does not lead to an oscillation around the true result, and therefore,
there is no obvious approach of reliably estimating prediction uncertainty of the computational model employed.
However, the result of a computational model is incomplete without an accurate uncertainty associated with it \cite{Irikura2004}.
Given a reliable uncertainty measure for a computational result, one could not only estimate the effects on observables derived from that result (through uncertainty propagation),
but also directly assess the quality of approximations in the model-development stage.
Finally, availability of prediction uncertainties would help select an appropriate computational model of sufficient accuracy for a problem at hand.

\clearpage

\section{Uncertainty Quantification in Computational Chemistry}

\subsection{Benchmark Studies}
It is generally assumed that performance statistics based on benchmark systems are good estimates for the prediction uncertainty of a quantum chemical method.
Due to the availability of large amounts of experimental and computational reference data (for a recent review see Ref.~\cite{Peverati2014}),
benchmark studies are carried out to provide statistical quantities such as the mean absolute error (MAE),
\begin{equation}
\text{MAE}_m = \frac{1}{N} \sum_{s=1}^{N} | e_{m,s} |,
\end{equation}
and the largest absolute error (LAE),
\begin{equation}
\text{LAE}_m = \text{max} \{|e_{m,1}|, |e_{m,2}|, ..., |e_{m,{N}}|\},
\end{equation}
with $e_{m,s} = c_{m,s} - o_s$ and $N$ being the size of the data set.
Here, the error $e_{m,s}$ of model $m$ with respect to system $s$ (typically a molecule) is defined as the difference between the calculated result $c_{m,s}$ and the experimental or computational reference $o_{s}$.
These summarizing statistics are then applied to estimate the prediction uncertainty of a method of choice for a system of interest.

However, there is a major caveat associated with this approach: the assumption that such statistics are transferable to a system not represented in the reference data set is generally invalid.
In Table~\ref{tab:maes} the MAE of common density functionals with respect to ligand dissociation energies of transition metal complexes from three previous studies are compared.
The WCCR10 data set \cite{Weymuth2014} consists of 10 ligand dissociation energies of large cationic transition metal complexes.
The 3dBE70 database \cite{Zhang2013} contains average bond energies of 70 transition metal compounds.
The data set by Furche and Perdew \cite{Furche2006} containing 18 dissociation energies of transition metal compounds is herein abbreviated as FP06.
\begin{table}
\centering
\caption{Mean absolute error (MAE$_m$) of ligand dissociation energies (kJ/mol) calculated with a selection of common density functionals $m$ taken from the literature.
}
\label{tab:maes}
\begin{tabular}{lrrrr}
\hline
Model $m$ &  WCCR10~\cite{Weymuth2014}   &  3dBE70~\cite{Zhang2013}  &  FP06~\cite{Furche2006} \\
\hline
B3LYP \cite{Lee1988,Becke1993,Becke1988}           &  39.1                     &  20.9                   &  50.2 \\
PBE \cite{Perdew1996a,Adamo1999,Perdew1996}        &  31.8                     &  25.5                   &  45.2 \\
TPSSh \cite{Staroverov2003}                        &  32.0                     &  17.6                   &  40.6 \\
\hline
\end{tabular}
\end{table}
The comparison of the different benchmark studies shows that the MAEs are strongly data set dependent.
For instance, the spread of MAEs ranges from 17.6 to 40.6~kJ/mol in the case of the TPSSh density functional.

Even for small systems such as metal dimers the reported statistics can vary.
For example, for the dissociation energy of metal dimers the study by Furche and Perdew \cite{Furche2006} and Schultz et al.\ \cite{Schultz2005a} report MAEs of 50.6 and 69.9~kJ/mol, respectively.
This finding is in accordance with many studies demonstrating that the accuracy of density functionals varies strongly with the chemical system \cite{Curtiss1997,Curtiss2000,Salomon2002,Zhao2004,Curtiss2005,Riley2007,Weymuth2014},
and therefore, undermining the transferability of such performance statistics.
In the case of density functional theory, this lack of transferability is particularly critical to studies on transition metals since most of the benchmark data sets include only small (unsaturated and therefore atypical) compounds (e.g., transition metal hydrides such as FeH).

In addition, it can be seen from Table~\ref{tab:maes} that all MAEs are considerably large (a result is said to be within chemical accuracy if the expected error is within $\approx 4.2$~kJ/mol).
For the WCCR10 and FP06 data sets LAEs are reported as well (e.g., 83.4 and 157.3~kJ/mol, respectively, for the B3LYP functional).
MAE and LAE of this size are unacceptable for studies in which accurate reaction energy are of high importance.
In the framework of conventional transition state theory, an error of 30 kJ/mol in the barrier height of an elementary reaction step results in a reaction rate that is off by a factor of $10^5$.

Lastly, it should be noted that the uncertainty within the (experimental and computational) reference data is generally not accounted for \cite{Pernot2015}.

\begin{figure}[!h]
\centering
\includegraphics[width=\textwidth]{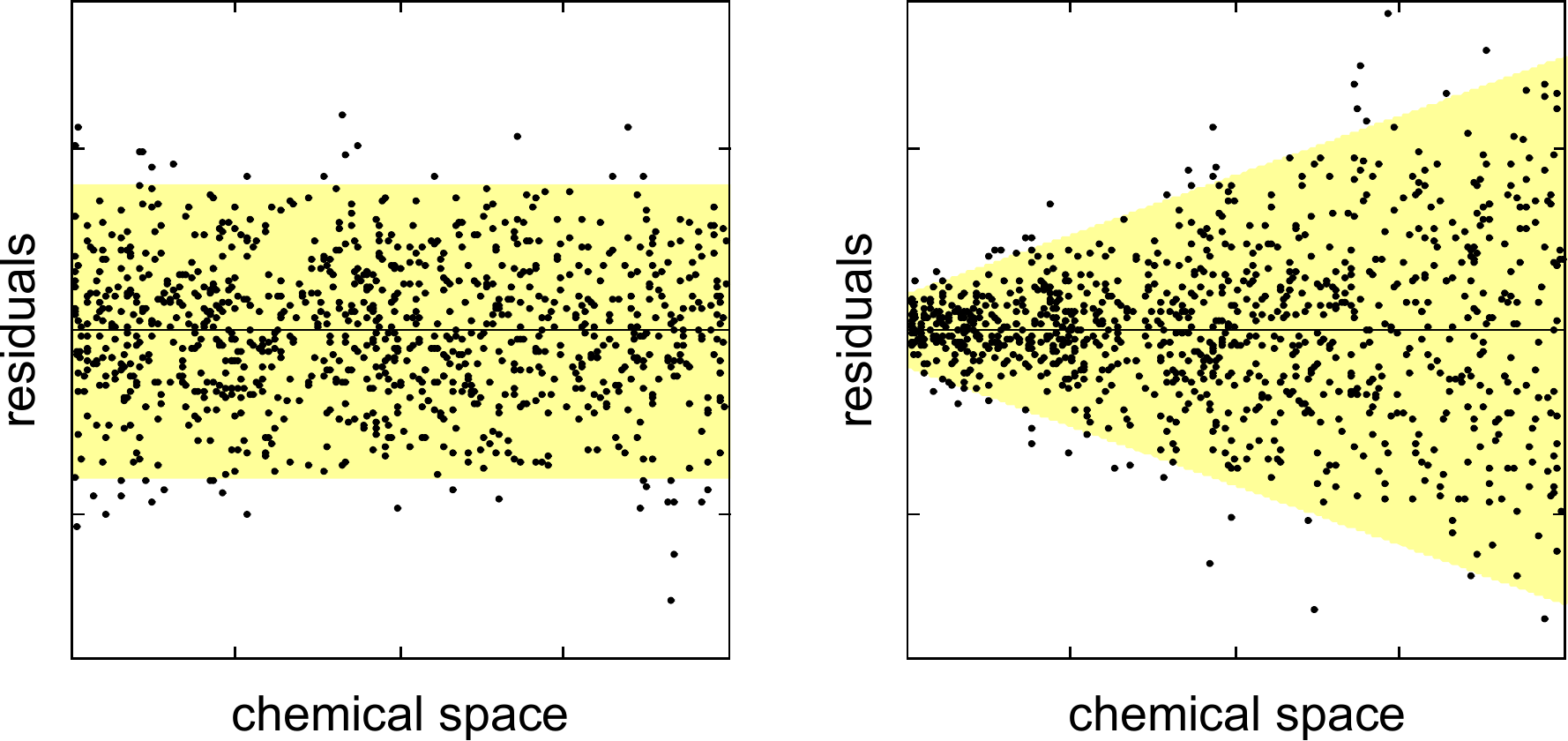}
\caption{Illustration of homoscedasticity (left) and heteroscedasticity (right) for synthetic data.
In the former case (left), the uncertainty (yellow 95\% confidence band) associated with an observable of interest is independent of the chemical system studied.
In the latter case (right), which is the more general case, the uncertainty associated with the observable of interest is a function of the chemical space.
The distance between two data points along the abscissa is thought to be inversely proportional to the similarity of the corresponding molecular structures.
Hence, if a prediction method is trained on a small hypervolume of the chemical space, it will not be possible to transfer the associated uncertainty to a larger hypervolume.
Moreover, since the variance function is generally unknown, also internal predictions (in the same hypervolume where the method has been trained) are unreliable.
}
\label{fig:variance}
\end{figure}

In Figure \ref{fig:variance}, we illustrate the system dependency of an arbitrary observable given an adequate computational model (see Section \ref{sec:inadequacy} for a definition of model adequacy).
The transferability of statistical measures such as the MAE would only be valid in the ideal case of homoscedasticity (Figure \ref{fig:variance}, left),
where the prediction uncertainty is independent of the input, here, chemical space (the space of all chemical compounds, e.g.\ molecules, where small distances indicate high structural similarity).

So far, there exists no strategy to develop approximate quantum chemical methods with system-invariant uncertainty (homoscedasticity), which is not to be confused with strategies to develop systematically improvable methods
(such as the coupled cluster expansion, which still reveals systematic errors due to the truncation of the degree of excitation --- even if the degree is taken to be rather high).
Consequently, we are generally faced with approximations yielding heteroscedastic results (Figure \ref{fig:variance}, right), where the prediction uncertainty somehow depends on the nature of the chemical system.

This dependency is generally unknown (not as indicated in the right frame of Figure \ref{fig:variance}), which also implies that estimation of prediction uncertainty for data lying in the same region of the chemical space employed for model training can be unreliable.
Noteworthy, the Hohenberg--Kohn functional would, in principle, yield results with system-invariant accuracy (for chemical systems in their electronic ground states),
however, this is not the case in practice due to the approximations of the exchange--correlation density functional.

\subsection{Reference Methods}
Due to the continuous advancement in accurate and efficient black-box methods (such as explicitly correlated coupled cluster theory, for a review see Ref.~\cite{Klopper2006}) and the increase of computational power,
it is believed that these gold standard methods will, eventually, become the standard method of choice.
In this case, uncertainty estimation will be less important if chemical accuracy is reached and considered sufficient.
For higher accuracy also standard coupled cluster models will require rigorous error estimation.
Although the system size for which these methods are feasible increases due to constant method-development efforts, less accurate methods are usually chosen for feasibility reasons when a large number of calculations must be carried out.
This is the case for extensive explorations of vast reaction networks \cite{Zimmerman2013,Zimmerman2015a,Rappoport2014,Zubarev2015,Bergeler2015}, screening studies \cite{Olivares-Amaya2011,Hachmann2011},
and reactive molecular dynamics simulations \cite{vanDuin2001,Dontgen2015,Saitta2014}.

\subsection{Error Assignment}
The identification and separation of sources of uncertainty is difficult, since multiple approximations of unequal accuracy are made during method development.
For example, in density functional theory, the exact density functional is approximated in a rather involved way.
In standard coupled cluster theory, the wave function is based on a single reference (Slater determinant).
On the one hand, these and other sources of uncertainty may combine in an arbitrary manner and even lead to counter-intuitive total errors \cite{Pernot2015a}.
For example, coincidental error compensation can lead to overestimation of prediction accuracy.
This is an effect often encountered in density functional theory.
For instance, the success of the B3LYP \cite{Lee1988,Becke1993,Becke1988} functional together with the poor 6-31G* basis set \cite{Hehre1972} is often attributed to error cancellation \cite{Kruse2012}.
Error compensation was also reported for coupled cluster methods, for instance, CCSD(T) was found to provide more accurate results than CCSDT in combination
with certain one-electron basis sets \cite{Feller2006}.
On the other hand, there are approximations (e.g., considering the atomic nucleus as a point charge rather than as an extended charge distribution, ignoring certain relativistic effects)
that are local (atomic) and cancel out for reaction energies (or valence properties).

In recent work \cite{Proppe2016}, we discussed the approximations necessary for the calculation of thermochemical properties in liquid phase.
We concluded that the contribution of each approximation to the overall error is difficult to determine but necessary for meaningful conclusions from subsequent analyses such as kinetic studies.

\section{Uncertainty Classification}
In general, one distinguishes between three main sources of uncertainty: parameter uncertainty, numerical uncertainty, and systematic errors due to inconsistent data and inadequate model approximations (here, to the fundamental theory of chemistry, QED) \cite{Pernot2016}.
Except for stochastic models (e.g., Monte Carlo simulations), numerical uncertainty is expected to be negligible and will not be discussed in the following.
The remaining sources of uncertainty are elaborated on and approaches for their remedy are elucidated.

\subsection{Parameter Uncertainty}

For the prediction of properties of chemical systems not included in the training of a computational model, one needs to estimate the uncertainty of its parameters in addition to their ``best'' values (obtained from minimizing a cost function such as the sum of squared residuals).
Otherwise, one would neglect a (potentially essential) component in determining the prediction uncertainty of a computational model.
Parameter uncertainty is a result of random and systematic errors in both the reference data and the computational model under consideration (see Section \ref{sec:inadequacy}), in particular if the number of reference data is small.
Only for a large number of data and a given domain of application (e.g., a specific volume of chemical space), parameter uncertainty becomes negligible.

Parameter uncertainty can be estimated in many ways, for example, through Bayesian inference \cite{Bishop2009} or through resampling methods such as bootstrapping \cite{Chernick1999}.
In the latter case, the reference data set itself replaces the critical assumption of a parametric population distribution underlying the data (for instance, the normal distribution is parameterized by mean and variance).

To obtain information on parameter uncertainty with bootstrapping, one draws as many data points as contained in the reference set, but \textit{with replacement}.
Every such bootstrap sample will yield different parameter values compared to the original (reference) sample, the ensemble of which allows estimation of parameter uncertainty.

Assuming that systematic errors in the computational model have been eliminated (for instance, by a posteriori corrections of its results \cite{Pernot2015}),
the effect of the reference set employed on the parameter distributions (and, as a consequence thereof, parameter uncertainty) remains to be examined.
If the reference data contain systematic errors, small changes in its composition (e.g., removal or addition of a few data points) may have a significant effect on mean, variance, and higher moments of the parameter distributions.
A well-established method for the detection of such data inconsistencies is the jackknife \cite{Chernick1999}, where changes in the parameter distributions are identified by removing individual data points.
Given a reference set comprising $N$ data points, one obtains $N$ jackknife estimates of the parameter distributions, each of them inferred from the reference set with the $s$-th data point removed ($s = 1,...,N$).

We combined the jackknife with bootstrapping to examine systematic data errors in the calibration of a prediction model for the $^{57}$Fe M\"ossbauer isomer shift \cite{Proppe2017}.
The corresponding theory \cite{Gutlich2011} postulates a linear relation with the electron density at the iron nucleus, which varies due to the chemical environment in which it is embedded.
We studied 44 iron complexes featuring high chemical diversity for which we calculated the electron contact density on the basis of 12 density functionals across Jacob's ladder (from local density approximations to meta-hybrid generalized gradient approximations).
We obtained 12 data sets with pairs of experimental isomer shifts and calculated electron contact densities, which only differ in the values of the latter quantity.
We identified an iron complex as potentially critical if its removal from the data set has a significant effect on the bootstrapped parameter distributions and, therefore, on the uncertainty of isomer shift predictions.
Noteworthy, four (chemically dissimilar) iron complexes were identified as potentially critical for all density functionals applied, which indicates either systematic experimental errors (hard to validate in hindsight) or unrepresentative molecular structures.

After removal of the critical data points in our M\"ossbauer study, we examined the effect of both composition and number of data on the ranking of density functionals, which is based on reliable prediction-uncertainty estimations.
For this purpose, we created 10,000 synthetical data sets of different size (from 5 to 1,000 data points) with the bootstrap approach.
We found that the density functional ranking is very sensitive to the specific data set when it comprises only 5 data points, still quite sensitive for 40 data points, and converges only for a large number of data points (1,000).
Our study \cite{Proppe2017} showed that conclusions about prediction uncertainty and rankings of computational models based on a single data set are sensitive to errors, and that bootstrapping is a simple and fast method to avoid them.

\subsection{Model Inadequacy}
\label{sec:inadequacy}
\begin{figure}[!h]
\centering
\includegraphics[width=.6\textwidth]{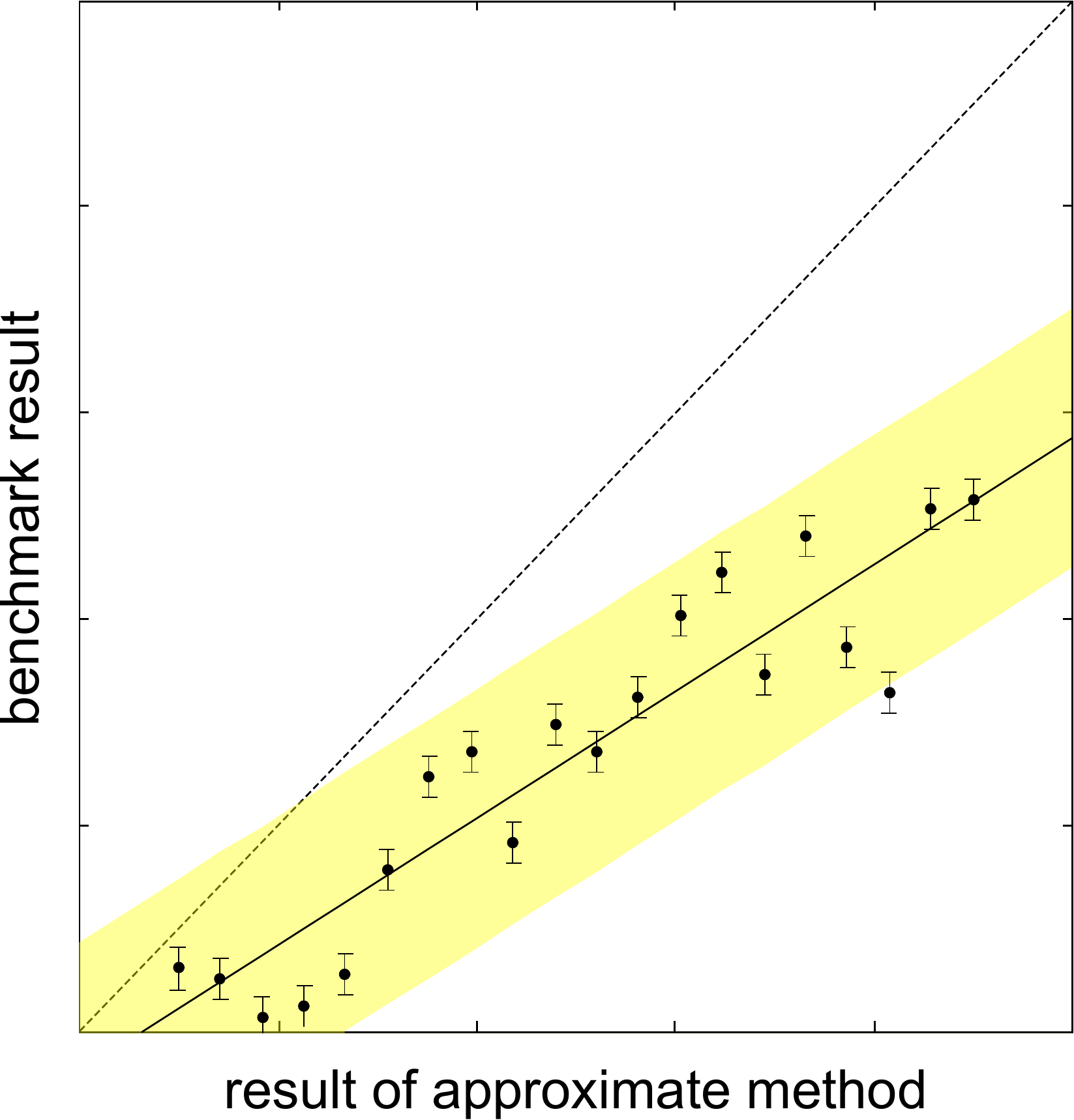}
\caption{Illustration of systematic and random model inadequacy for synthetic data.
For an adequate approximate method, the data would scatter around the line through the origin (dashed line).
Here, however, the results of the approximate method reveal a non-constant deviation from the benchmark results (obtained from measurements or very accurate calculations), which is an indication of systematic model inadequacy.
An a posteriori correction of the approximate method can be realized by fitting a linear calibration function to the data (solid line).
The scatter of data around the calibration line appears to be random, but the residuals are on average significantly larger than the uncertainty in the benchmark results (indicated by error bars representing two standard deviations).
This effect is referred to as random model inadequacy \cite{Pernot2015} and implies that the uncertainty of the approximate method (represented by the yellow 95\% prediction band) exceeds the uncertainty of the benchmark.
Here, the error bars are obviously narrower than the prediction band.
}
\label{fig:calibration}
\end{figure}

An inadequate computational model is not able to reproduce reference data within their uncertainty range \cite{Pernot2016}, i.e., the model under- or overestimates the uncertainty of the reference data.
Underestimating prediction uncertainty is a result of overfitting, where the computational model is too flexible (features too many parameters) such that it does not only fit the explainable part of the reference data (the underlying physics), but also its unexplainable part (noise).
By contrast, underfitting is caused by models which are too rigid (possess too few parameters) to fit the explainable part of the reference data, leading to overestimation of prediction uncertainty.
Moreover, model inadequacy can be divided into an explainable (systematic) and an unexplainable (random) part, which is illustrated in Figure \ref{fig:calibration}.

For instance, most quantum chemical methods (with the exception of multi-configurational methods) struggle to correctly describe two hydrogen atoms at large distance.
In fact, all density functionals fail to describe stretched H$_2^+$ and H$_2$ \cite{Cohen2012}.
The smoothness of the corresponding energy--distance plots (see, for instance, Figure~2 in Ref.~\cite{Cohen2012}) reveals that random model inadequacy plays a negligible role in this ``simple'' case of two nuclei.
However, the fact that all of these energy--distance plots reveal a non-constant deviation from those obtained with accurate multi-configurational methods shows high significance of systematic model inadequacy.
While in this special case, model inadequacy could be easily eliminated by fitting a reasonable function linking data from benchmark and approximate calculations, the situation will become much more complicated if a larger fraction of chemical space is considered.
For instance, due to their complex electronic structure, molecular structures containing transition metals are challenging targets for current quantum chemical methods.
Despite containing adjustable empirical parameters, many density functionals fail to achieve a statistically valid description of these systems \cite{Weymuth2015}.
We showed, for example, that the parameters of a standard functional are flexible enough to be chosen to exactly reproduce each coordination energies of a data set containing large organometallic complexes \cite{Weymuth2015}.
However, due to model inadequacy, there exists no unique parameter set that is equally accurate for all coordination energies in this data set at the same time.

Note that model inadequacy is difficult to distinguish from data inconsistency.
If the reference data contain systematic errors, even high-accuracy models would not be able to reproduce the reference data.
In that case, it would be the wrong decision to improve on the computational model (high overfitting tendency).
We showed at the example of M\"ossbauer isomer shift prediction \cite{Proppe2017} that application of the jackknife combined with bootstrapping  on a diverse selection of model approximations (see introduction to Section \ref{sec:inadequacy}) supports unraveling the two effects (data inconsistency and model inadequacy).

Given the reference data is corrected for inconsistencies, there are several tools at hand to tackle model inadequacy \cite{Pernot2016, Pernot2016a}:
one can improve the underlying model, reduce the domain of application, or correct predictions through a statistical calibration approach.

\subsubsection{Model Improvement}
If the computational model at hand is systematically improvable (as, for instance, in the case of a coupled cluster expansion) reduction of model inadequacy is, in principle, straightforward.
However, such methods are currently limited to relatively small system sizes and a few structures to be considered.

In density functional theory, model improvement is often referred to as climbing up Jacob's ladder \cite{Perdew2001}.
Higher rungs incorporate increasingly complex ingredients constructed from the density or the Kohn--Sham orbitals (e.g., gradient and Laplacian of the electron density, kinetic energy density).
The original proposition of a ladder is that each rung satisfies certain exact constraints (there exist 17 of them, see the Supplementary Material of Ref.~\cite{Sun2015}) and the next higher rung should be based on the previous rungs \cite{Cohen2012}.
Since the exact density functional is not known and the number of known exact constraints is severely limited, systematic model improvement is not trivial.

In fact, a very recent study has shown that current developments steer away from systematic model improvement and towards functionals of empirical nature lacking physical rigor \cite{Medvedev2017}.
Most density functional development is focused on energies, implicitly assuming that functionals producing better energies become better approximations of the exact functional.
The exact functional will produce the correct energy only if the input electron density is exact as well.
By contrast, Peverati and Truhlar \cite{Peverati2014} argued that exact constraints can be neglected for the sake of greater flexibility in the energy fitting.
However, such flexibility comes at the cost of reduced transferability (due to overfitting, cf.\ introduction to Section \ref{sec:inadequacy}) to both other observables and chemical systems not included in the training of the computational model.
To avoid loss of model transferability, Mardirossian and Head-Gordon suggest a validation approach where the performance of a certain density functional is assessed for a data set not involved in the training of that density functional \cite{Mardirossian2014, Mardirossian2015}.
This way, one can successively increase model flexibility until the validation indicates a decrease of transferability (due to an increase in the performance statistics chosen).

Composite methods such as Gaussian-$n$ (G-$n$) \cite{Pople1989,Curtiss1991,Curtiss1998,Curtiss2007}, Weizmann (W-$n$) \cite{Martin1999,Boese2004,Karton2006}, and HEAT \cite{Tajti2004} aim for high accuracy by combining the results of several calculations.
They build a hierarchy of computational thermochemistry methods which allows the calculation of molecular properties such as total atomization energies and heats of formation to a high accuracy.
The W-4 method calculated atomization energies of a set of small molecules with an MAE below 1~kJ/mol \cite{Karton2006}.
Similarly, the HEAT protocol predicted enthalpies of formation with an accuracy below 1~kJ/mol for 31 atoms and small molecules \cite{Tajti2004}.
These protocols rely on computationally expensive coupled cluster calculations including high excitations.
The HEAT method applies additional calculations (e.g., the diagonal Born–Oppenheimer correction) to be able to reproduce experimental results to higher accuracy.
While the results from such methods are promising, the computational cost is far too high for large-scale applications mentioned above.

Errors in estimating prediction uncertainty due to model inadequacy can be eliminated not only by \textit{internal} correction of a computational model (see the examples above),
but also through \textit{external} correction of the results produced with a computational model \cite{Pernot2015}.
The simplest external corrections are linear functions, which are applied in the prediction of, for example, vibrational frequencies \cite{Rauhut1995, Neugebauer2003, Irikura2005} or M\"ossbauer isomer shifts \cite{Neese2002, Han2006, Romelt2009, Bochevarov2010, Gubler2013, Proppe2017}.
In such cases, parameter inference (calibration) can be much more efficient than internal calibration of the result-generating model.
A drawback is the loss of transferability to other observables since the external calibration model corrects an expectation value of a certain observable and not its underlying wave function, which is the unique common physical ground of all observables.

\subsubsection{Reduction of Domain of Application}

\begin{figure}[!h]
\centering
\includegraphics[width=.6\textwidth]{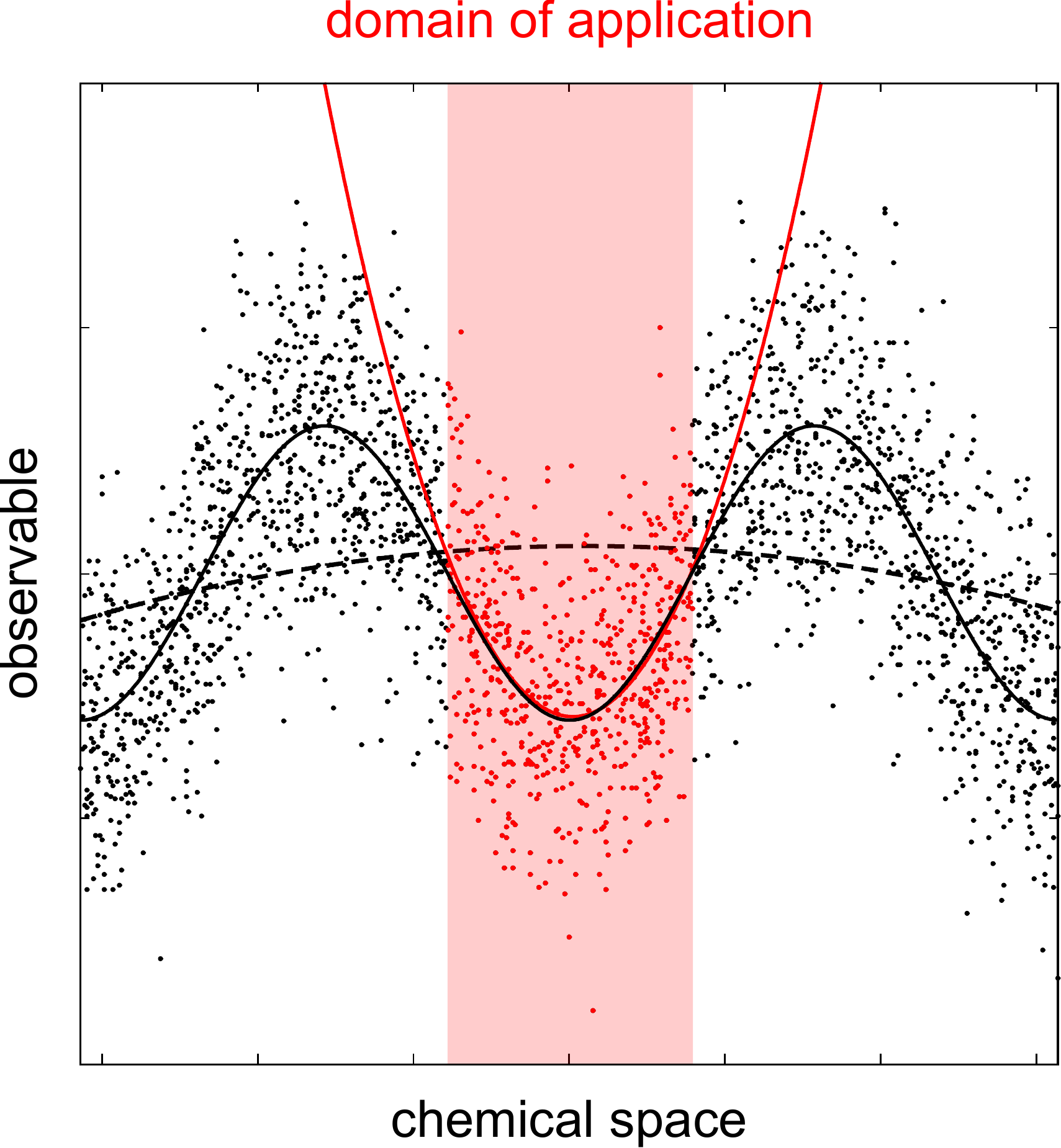}
\caption{Illustration of model inadequacy for synthetic data.
The black solid curve (cosine) is set to be the correct model (no under- or overfitting) to the entire domain of application.
The distance between two data points along the abscissa is thought to be inversely proportional to the similarity of the corresponding molecular structures.
We assume that our approximate model is a quadratic function.
If our reference data (dots) are spread across the entire domain of chemical space shown, we will observe a systematic deviation of the observable from our approximate model (dashed line).
However, if we choose a specific domain of application (red shaded area), our approximate model (red curve) will be a good approximation to the correct model.
To avoid model inadequacy in this case, we can either improve our model by increasing its complexity (here, to a cosine) or reduce the domain of application (to the red region).
}
\label{fig:inadequacy}
\end{figure}

Another way of reducing model inadequacy is by training a computational model on a smaller domain of chemical space \cite{vonLilienfeld2013}, i.e.\ a small set of similar molecules such as sugars or amino acids.
For example, due to the strong approximations made during method development (to gain efficiency), semi-empirical methods exhibit model inadequacy,
which they attempt to remedy by introducing parameters which are then fitted to a specific data set (for a recent review see Ref.~\cite{Thiel2014}).
This data set comprises a selection of molecules for which the resulting method is tailored.
In fact, semi-empirical methods have been reparameterized to improve their description of a single molecule \cite{Wu2013}.
Similarly, density functionals were developed for specific applications, e.g., for kinetic studies \cite{Lynch2000,Proppe2016}.
In Figure \ref{fig:inadequacy}, the effect of the domain of application on model inadequacy is illustrated by a toy model.

We applied the domain-reduction approach for the development of a system-specific density functional that was derived on a sound physical basis \cite{Simm2016}.
We re-parameterized a range-separated hybrid functional to reproduce (computational) energy differences between isomers of a transition-metal catalyst, which refers to a small volume of chemical space (cf.\ Figure \ref{fig:inadequacy}).
While the resulting functional turned out to be more accurate than any popular density functional and the error estimates were in reasonable accordance with the residuals, the effect of model inadequacy prevailed to a certain degree.
The functional is unable to describe the complex electronic structure of the transition metal complexes in selected cases.

\subsubsection{Increase of Parameter Uncertainty}

One can attempt to compensate model inadequacy by a controlled increase in parameter uncertainty.
This way, one can build a statistical method with prediction uncertainty representative of the model residuals (deviation of benchmark data from model predictions).

In 2005, N{\o}rskov, Sethna, Jacobsen, and co-workers implemented this approach for error estimation of results from density functionals \cite{Mortensen2005} (see also Refs.\ \cite{Brown2003, Frederiksen2004, Petzold2012}).
Instead of considering only the best-fit parameters of a density functional, they assigned a conditional probability distribution to them so that a mean and a variance can be assigned to each computational result.
While promising general-purpose non-hybrid density functionals were designed within this framework (e.g.\ BEEF-vdW \cite{Wellendorff2012} and mBEEF \cite{Wellendorff2014, Pandey2015}),
the accuracy of uncertainty predictions remains unsatisfying \cite{Pernot2016}.
This limitation can be attributed to model inadequacy and the heteroscedasticity of the large domain of chemical space to which they applied the functionals.

Compared to improving the computational model itself, increasing parameter uncertainty is straightforward as it only requires modification of the unknown part (parameter distributions) of an otherwise known model.
Compared to external calibration (a posteriori correction of results obtained from a computational model), increasing parameter uncertainty in the corresponding prediction model preserves its transferability to other observables than the reference observable (for which model inadequacy has been corrected).
While increased parameter uncertainty seems to be clearly favorable over model improvement when it comes to reliably estimating prediction uncertainty for any observable obtained on the basis of a given computational model, it does not resolve the issue of model inadequacy per se.
For instance, in multiscale modeling where the target observable is built on a hierarchy of other observables with decreasing time and/or length scales, all uncertainties inferred at low levels (small time/length scales) will propagate to the final prediction uncertainty (see Section \ref{sec:propagation}).
Consequently, increasing parameter uncertainty at low levels can lead to a prediction uncertainty so large that no sensible conclusions can be drawn from it.

Recently, we demonstrated the sensitivity of final prediction uncertainty in multiscale modeling for the inference of kinetic reaction networks based on quantum-chemical methods \cite{Proppe2016}.
Uncertainty in the electronic energy propagates to all energy contributions based on nuclear motion, to any kind of free energy, to rate constants, and to concentration fluxes of chemical species (an incomplete but lucid list).
The dependencies between these observables are partially exponential, which requires the minimization of systematic errors in the low-level observables (instead of hiding them in increased parameter uncertainty).
In such cases, the only possible way to obtain reasonably small prediction uncertainties is the systematic improvement of methods on the different length and time scales.

\subsection{Uncertainty propagation}
\label{sec:propagation}

Uncertainty propagation describes the process of transferring the uncertainty of model parameters to the uncertainty of model predictions.
Prediction uncertainty can be assessed through calibration against reference data.
There are two types of calibration: \textit{internal} calibration of a computational model (adjustment of method-inherent parameters such as those of a density functional) and \textit{external} calibration of the results produced with a computational model \cite{Pernot2015}.
If further calibration is not necessary (in the ideal case when systematic errors are absent), the observation (reference value) $o_s$ of a system $s$ including \textit{known} uncertainty $u_s$ is completely determined by the essentially variance-free result $c_{m,s}$ of a computational model $m$ plus random error $\varepsilon_{m,s}$ (drawn from a zero-mean distribution with variance $u_s^2$),

\begin{equation}
\label{eq:prediction}
o_s = c_{m,s} + \varepsilon_{m,s} \, .
\end{equation}

In internal calibration, the method-inherent parameters need to be adjusted such that Eq.\ (\ref{eq:prediction}) is fulfilled, which requires a functional form with sufficient flexibility.
In external calibration, we expand the expression on the right-hand side of Eq.\ (\ref{eq:prediction}) by building a calibration model $f(c_{m,s},\mathbf{w}_m)$ around the computed results,

\begin{equation}
o_s = f(c_{m,s},\mathbf{w}_m) + \varepsilon_{m,s} \, ,
\end{equation}

where $\mathbf{w}_m$ is the vector of parameters of the external calibration model.

To determine the uncertainty of a virtual measurement (prediction), $u(c_{m,s})$, on the basis of the computed result $c_{m,s}$ for a physical measurement \textit{not} included in the training data set, we need to propagate the uncertainty of $\mathbf{w}_m$ to that of $f(c_{m,s},\mathbf{w}_m)$.
The simplest way to do so is \textit{linear uncertainty propagation}, where the uncertainty of the external calibration model is approximated by its first partial derivative with respect to its parameters,

\begin{equation}
\label{eq:propagation}
u(c_{m,s})^2 = \sum_{ij}\frac{\partial f(c_{m,s},\mathbf{w}_m)}{\partial w_{i,m}}\frac{\partial f(c_{m,s},\mathbf{w}_m)}{\partial w_{j,m}} \mathbb{V}[w_{i,m}, w_{j,m}] \, ,
\end{equation}

where $i = 1,...,M$ and $j = 1,...,M$ index the $M$ parameters contained in $\mathbf{w}_m$, and $\mathbb{V}[w_{i,m}, w_{j,m}]$ is the $ij$-th element of the covariance matrix of the parameters.
When changing the $f(c_{m,s},\mathbf{w}_m)$ terms in Eq.\ (\ref{eq:propagation}) to $c_{m,s}$, we obtain an expression for linear uncertainty propagation in the case of internal calibration, where $\mathbf{w}_m$ now represents the parameters of the computational model.

If the calibration model is linear in the parameters, linear uncertainty propagation is an exact procedure.
For calibration models being nonlinear in the parameters, higher derivatives of the calibration model may become necessary, the calculation of which is often unfeasible.
In those cases, stochastic methods such as Monte Carlo uncertainty propagation are applied \cite{Ogilvie1984}.

\section{Conclusions}
We argued that a procedure for quantifying the uncertainty associated with computational models, in particular with quantum chemical calculations, is mandatory despite their first-principles character.
Otherwise, it may be difficult to draw meaningful conclusions.
Unfortunately, this procedure is neither well established nor straightforward.
The abundance of benchmark studies reporting (potentially misleading) statistical measures such as the MAE and LAE,
the hope for accurate post-Hartree--Fock methods to become routinely and universally applicable, and the difficulty of identifying the source of error,
largely prevented the development of novel approaches for reliable error estimation.

We illustrated the different sources of errors and how to tackle them.
We stress that a clear differentiation between the different sources of error is critical for the effective application of countermeasures.
While numerical errors can often be controlled, model inadequacy and parameter uncertainty remain a major issue in quantum chemistry.
Reducing model inadequacy through model improvement is a popular approach, although not straightforward for most methods.
In these cases, statistical methods need to be applied in a rigorous way.
While in most cases this does not improve accuracy, it allows for reliable uncertainty predictions which are critical, especially if the error is propagated to subsequent investigations such as kinetic studies.

\section*{Acknowledgments}
This work has been financially supported by the Schweizerischer Nationalfonds (Project No. 200020\_169120).
GNS gratefully acknowledges support by a PhD fellowship of the Fonds der Chemischen Industrie.

%

\end{document}